\documentstyle[aas2pp4,amssym,psfig]{article}

\righthead{NIR Standard Stars}
\lefthead{Hunt et al.}

\begin{document}

\title{ Northern $JHK$ Standard Stars for Array Detectors}  

\author{ L. K. Hunt, F. Mannucci } 
\affil{ C. A. I. S. M. I. - C. N. R. }
\authoraddr{ Largo E. Fermi 5, I-50125 Firenze, Italy\\
Electronic mail: hunt@arcetri.astro.it; filippo@arcetri.astro.it}

\author { L. Testi\altaffilmark{1}, S. Migliorini, R. M. Stanga  } 
\affil{ Dipartimento di Astronomia, Universit\`a di Firenze}
\authoraddr{ Largo E. Fermi 5, I-50125 Firenze, Italy}

\and

\author { C. Baffa, F. Lisi, L. Vanzi\altaffilmark{2}}
\affil{ Osservatorio Astrofisico di Arcetri} 
\authoraddr{ Largo E. Fermi 5, I-50125 Firenze, Italy}

\altaffiltext{1}{Current Address: CalTech, Pasadena, CA, USA}
\altaffiltext{2}{Current Address: CAISMI-CNR, Firenze, Italy}

\begin{abstract}
We report $J$, $H$, and $K$ photometry
of 86 stars in 40 fields in the northern hemisphere.
The fields are smaller than or comparable to a 4$\times$4~arcmin
field-of-view, and are roughly uniformly distributed over the sky,
making them suitable for a homogeneous broadband calibration 
network for near-infrared panoramic detectors. 
$K$ magnitudes range from 8.5 to 14, and $J-K$ colors from -0.1 to 1.2.
The photometry is derived from a total of 3899 reduced images;
each star has been measured, on average, 26.0 times per filter on 5.5 nights.
Typical errors on the photometry are $\sim$~0\fm012.
\end{abstract}

\keywords{methods: data analysis --- techniques: photometric }

\section{Introduction} 

The widespread availability of near-infrared (NIR) panoramic detectors has
rendered possible many scientific programs which were unfeasible with
single-element photometers.
New and more sophisticated data acquisition and
reduction techniques have successfully exploited the capabilities of
the new technology while, at the same time, photometric calibration
has generally relied on older pre-existing standard networks based
on ``one-pixel'' photometry. 
Such networks include the SAAO system of Glass (\cite{glas}),
expanded and rationalized by Carter (\cite{carter90}, \cite{carter95}), comprising
probably the most comprehensive and best-observed list available;
the MSO system defined by Jones \& Hyland (\cite{jone80}, \cite{jone82}),
now more or less supplanted or absorbed into the AAO system (Allen \&
Cragg \cite{alle});
the ESO system (Engels \cite{enge}; Bouchet, Schmider, \& Manfroid \cite{bouchet});
and the system which appears to the greatest extent to have inherited
the original mantle of the photometry of H. L. Johnson in the 1960s,
the CIT (Caltech/Cerro-Tololo) system of Frogel et al. (\cite{frog})
updated by Elias et al. (\cite{elia82}).
This last forms the basis of the unpublished, but widely used,
hybrid standard star set maintained at the 3.8~m UK Infrared Telescope
(UKIRT).

All these comprise relatively bright stars suitable for photometry 
at small- and medium-sized telescopes with instruments which do not have the
limited well capacities of array elements.
However, when observed with array detectors even on moderate-sized
telescopes, stars with  K $\lesssim$ 7.5 produce saturated pixels
unless defocussed or observed in non-standard modes with extremely short
on-chip integration times,
neither of which stratagem is conducive to precise and homogeneous calibration.

The ``UKIRT Faint Standards'' (Casali \& Ha-warden  \cite{casa}) upon
which the calibration of this work is based comprise a new set of much
fainter stars chosen and observed at UKIRT to facilitate observations
with panoramic detectors with limited dynamic range.
However, they are relatively few in number and isolated, occur mostly around
the celestial equator, and many of the stars are too faint 
to be useful for small- or medium-sized telescopes.
Only preliminary results are currently available for the UKIRT Faint Standards,
although this situation is actively being remedied by the expansion
and reobservation of the list at UKIRT.

We present here a set of
$J$, $H$, and $K$ photometric measurements, obtained with the Arcetri
NICMOS3 camera, ARNICA. 
The photometry
comprises 86 stars in 40 fields observable from the northern hemisphere.
The selection of the standard fields is described in Section~2, followed by
a discussion of observing and data reduction techniques in Section~3.
Section 4 presents the photometry and a comparison with other photometric systems. 

\section{The Sample Fields} 

The sample was designed around the capabilities of the large-format NICMOS
and InSb arrays mounted in cameras at small- to medium-sized telescopes.
In particular, stars with relatively faint K magnitudes were required 
(8.5$\lesssim$~K~$\lesssim$14.0), with as wide a color range as possible
(-0.2 $\lesssim$~J--K~$\gtrsim$1.0).
Moreover, the calibrated stars should be situated in relatively
uncrowded fields so as to avoid confusion and facilitate field
identification. 
At the same time, more than one calibrated star should be available
in a 2$\times$2~arcmin field-of-view (FOV).
Finally,
the range in apparent position was chosen to provide 
good sky coverage for northern hemisphere sites with latitudes between
35 and 45$^\circ$. 

To facilitate initial calibration and minimize the effects of color
terms, we selected a set of 15 SAO stars 
with spectral type A0 and V magnitudes of roughly 9.
These SAO stars are distributed uniformly in R.A and about  
a declination of 40$^\circ$.
The similarity of A-star NIR and visual magnitudes 
makes possible an instantaneous initial rough calibration 
and provides easy consistency checks.  
A stars have the further advantage that they tend not to be variable,
and, notwithstanding their relatively faint infrared magnitudes,
are easily visible at the telescope. 

Because we wanted to exploit the two-dimen-sional capability of 
panoramic detectors, we added to the list fields 
taken from the CCD standard stars given in Christian et al.
(\cite{chri}).
These fields comprise a large range in optical magnitude,
are minimally crowded, and are observable with a 
2$\times$2~arcmin FOV. 
All of the fields are in the vicinity of star clusters.

Stars selected from the UKIRT faint standard system 
(Casali \& Hawarden \cite{casa}) were used as primary calibration sources.
We extracted from the UKIRT list of faint standards those stars
with declinations $\ge\:0^\circ$ and with K magnitudes $\gtrsim$ 8.5;
$J-K$ colors for this subset range from $-0.2$ to $0.6$.
The UKIRT list comprises optical standards from
``selected areas'' (Landolt \cite{land}),
HST calibration objects (Turnshek et al. \cite{turn}), and stars in M67
(Eggen \& Sandage \cite{egge}) and in the globular clusters M3 (NGC\,5272)
and M13 (NGC\,6205). 
Stars surrounding the nominal calibration stars
in the UKIRT fields were used as program objects, and processed
in the same way as the A-star and CCD fields mentioned above.

Our goal was to calibrate as many stars as possible within the
central FOV of the camera observations.
Many of the central stars proved to be relatively isolated, so that
the mean number of stars per field is $\gtrsim\,$2.
The list of fields with coordinates of the individual stars for
which we obtained photometry is given in Table \ref{table:dat}.
The coordinates were determined from the Digitized Sky Survey\footnote{The
images on these disks are based on photographic data obtained using the 
Oschin Schmidt Telescope on Palomar Mountain.
The Palomar Observatory Sky Survey was funded by the National Geographic
Society.
The Oschin Schmidt Telescope is operated by the California Institute of
Technology and Palomar Observatory.
The plates were processed into the present compressed digital form with
their permission.
The {\it Digitized Sky Survey} was produced at the Space Telescope
Science Institute under U. S. Government grant NAG W--2166.}.
$K$-band finding charts for the 40 fields are illustrated in Fig. \ref{fig:charts}.

\placefigure{fig:charts}
\figcaption[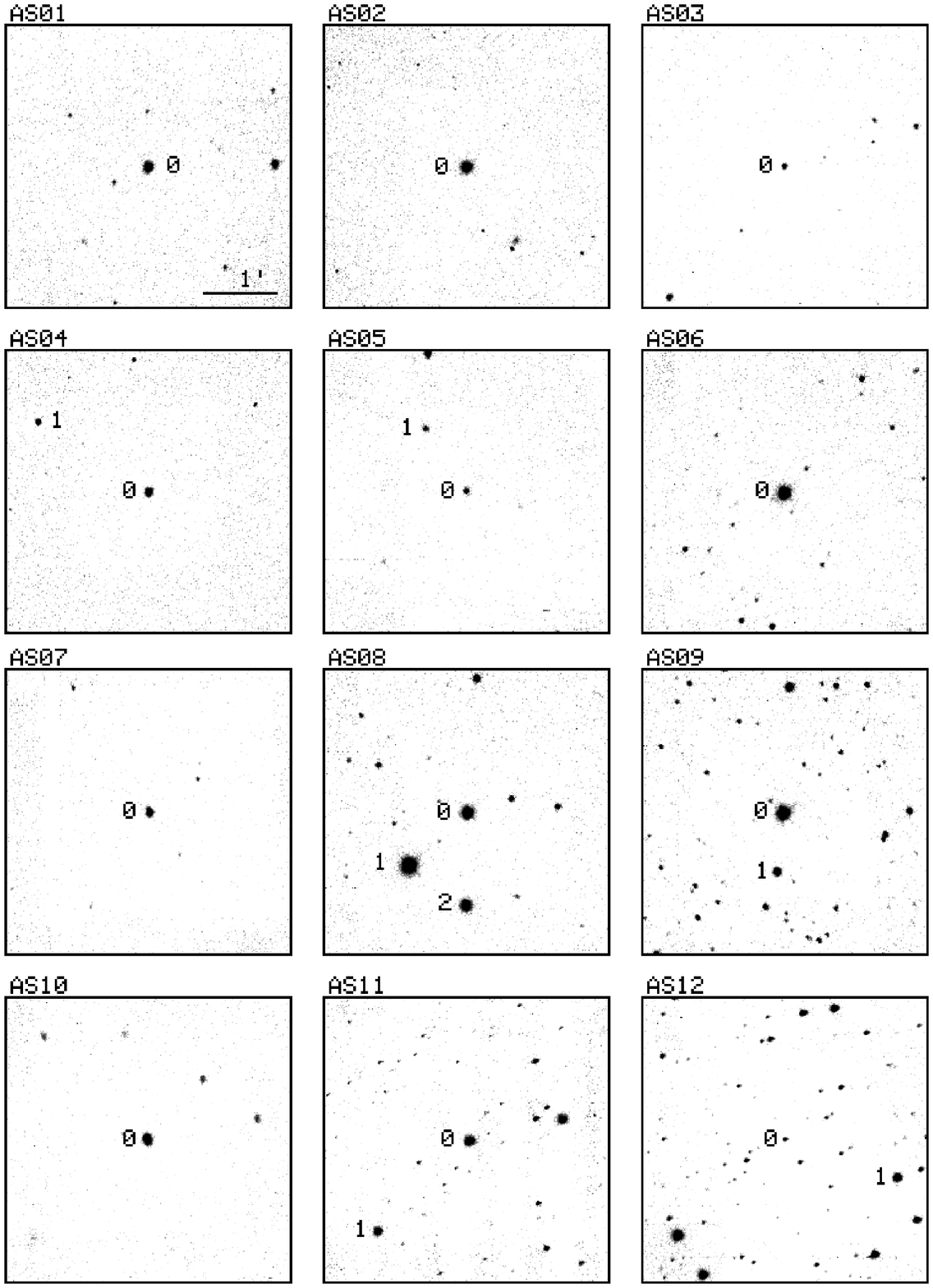]{ 
$K$-band finding charts. 
The field-of-view (FOV) is roughly 4$\times$4\,arcmin, and a 1\,arcmin segment
is shown in the upper-left panel in each page of the figure;
AS~39 has a smaller FOV, but the same scale, as suggested by the white border.
The finding charts are oriented with N up, and E to the left.
Central stars are labelled with a ``0'', and field stars with arbitrary 
sequential numbers, as also given in Table \ref{table:dat}.
\label{fig:charts} }

\bigskip

\section{Observations and Data Treatment} 

$J$ (1.2\micron ), $H$ (1.6\micron ), and $K$ (2.2 \micron) broadband images
of the standard star fields were acquired
with the Arcetri NIR camera, ARNICA, mounted at three telescopes:
1)~the 1.5-m f/20 Gornergrat Infrared Telescope\footnote{The TIRGO
(Gornergrat, Switzerland) is operated by CAISMI-CNR, Arcetri,
Firenze.} (TIRGO);
2)~the 2.56-m f/11 Nordic Optical Telescope\footnote{The NOT 
is operated on the island of La Palma jointly by 
Denmark, Finland, Norway, Sweden, in the
Spanish Observatorio del Roque de los Muchachos of the Instituto de
Astrofisica de Canarias.} (NOT);
3)~the 1.8-m f/9.5 Vatican Advanced Technology Telescope\footnote{The VATT 
(Mount Graham, Arizona) is composed of the Alice P. Lennon
Telescope and the Thomas J. Bannan Astrophysics Facility, both operated by 
the University of Arizona (Steward Observatory) and the Vatican Observatory
Research Group.} (VATT).
Data were obtained on 31 photometric nights in six runs from December, 1992, to April, 1997;
only 10 of these nights were dedicated exclusively to this project.

The camera, ARNICA, relies on a NICMOS3 256$\times$256 1--2.5~$\mu$m
HgCdTe scientific grade array,
and provides a 4$\times$4~arcmin FOV with 1~arcsec pixels at the TIRGO,
and 2$\times$2~arcmin with 0.5~arcsec pixels at the VATT and the NOT.
Details of the camera design and implementation are given
in Lisi et al. (\cite{lisi93}, \cite{lisi96}), 
and of the characterization of the
detector and camera performance in Hunt et al. (\cite{hunta}, \cite{hunt96}).

For each observation with each filter,
the center of the field was placed in five different positions
on the array: the first near the center, and the remaining four
in the center of each of the four quadrants. 
For $J$ and $H$, sky frames were used as the flat field correction.
To account for telescope emissivity at 2\,$\mu$m,
$K$-band observations were instead reduced using differential flat fields,
obtained independently.
Tests showed that these procedures minimize the residual (after flat-field
correction) spatial variation in the photometry (see Appendix).
All the data reduction was carried out with the ARNICA data reduction
package (Hunt et al. \cite{huntc}) in the IRAF\footnote{IRAF is the Image 
Analysis and Reduction Facility 
made available to the astronomical community by the National Optical
Astronomy Observatories, 
which are operated by AURA, Inc., under
contract with the U.S. National Science Foundation.} 
environment.

The images acquired in the first two (commissioning) runs --
about one-third of the database --
showed a residual spatial variation in the photometry
that was subsequently rectified by a modification in the camera optics.
To correct for this effect in those runs, 
an empirical second-order flat-field correction was applied to each image 
as described in Hunt et al. (\cite{huntb}, \cite{hunt96}).
We have checked that
this procedure produces no systematic trends relative to the later data.

\subsection{Photometry} 

Virtual aperture photometry was performed for each of 
the different positions separately.
The technique of moving the telescope to place the star
in different positions on the array implies that five measurements are available
for the central stars, but that field stars may have fewer measurements because
they may fall outside the array.
Note that we used relatively large apertures in order to encompass possible 
faint stars within 10\,arcsec of the program stars.
Images with discrepant values for the photometry were checked to evaluate
effects of any bad pixels, and these were either corrected, when possible, or
eliminated from further consideration.

Observations of the UKIRT faint standards throughout the night were
used to determine the nightly ``zero point'' (zp) for each band.
From five to eight stars from the UKIRT Faint Standard list were 
observed each night together with the program stars.
Most of the program star observations were obtained at 1.6 airmasses or less,
and every attempt was made to observe the calibrating stars over the
same range in airmass as the program stars.
Atmospheric extinction was corrected for by fitting simultaneously,
for each band for each run, an extinction coefficient common to all nights
and a zp that varied from night to night.
Typical extinction coefficients are $\sim\,-0.06$, $0.0$, and $-0.04$ mag/airmass
at $J$, $H$, and $K$, respectively.

In total, we calibrated 86 stars in 40 fields in 
$J$, $H$, and $K$ filters; the photometry is reported in 
Table~\ref{table:dat}.
The final values for the photometry
of the program stars were determined by combining,
in a weighted average, all the measurements, corrected for extinction,
from the different nights of observation. 
The errors given in Table \ref{table:dat} are the errors on the mean.
The exact prescription for these calculations is given in the Appendix.
For a few stars (e.g., AS16-1, AS39-1), 
the errors in one band are much smaller than the errors in the
remaining bands; these small errors probably result from
an observed scatter which is fortuitously small, and thus
the larger errors in the remaining bands
may be a more realistic estimate of the quality.
We note that the colors in the table are merely the differences of
the mean magnitudes, as we determined the photometry in each filter band
separately, instead of giving precedence to one band and determining
mean colors.

\section{Results and Discussion}

\begin{figure}
\plotone{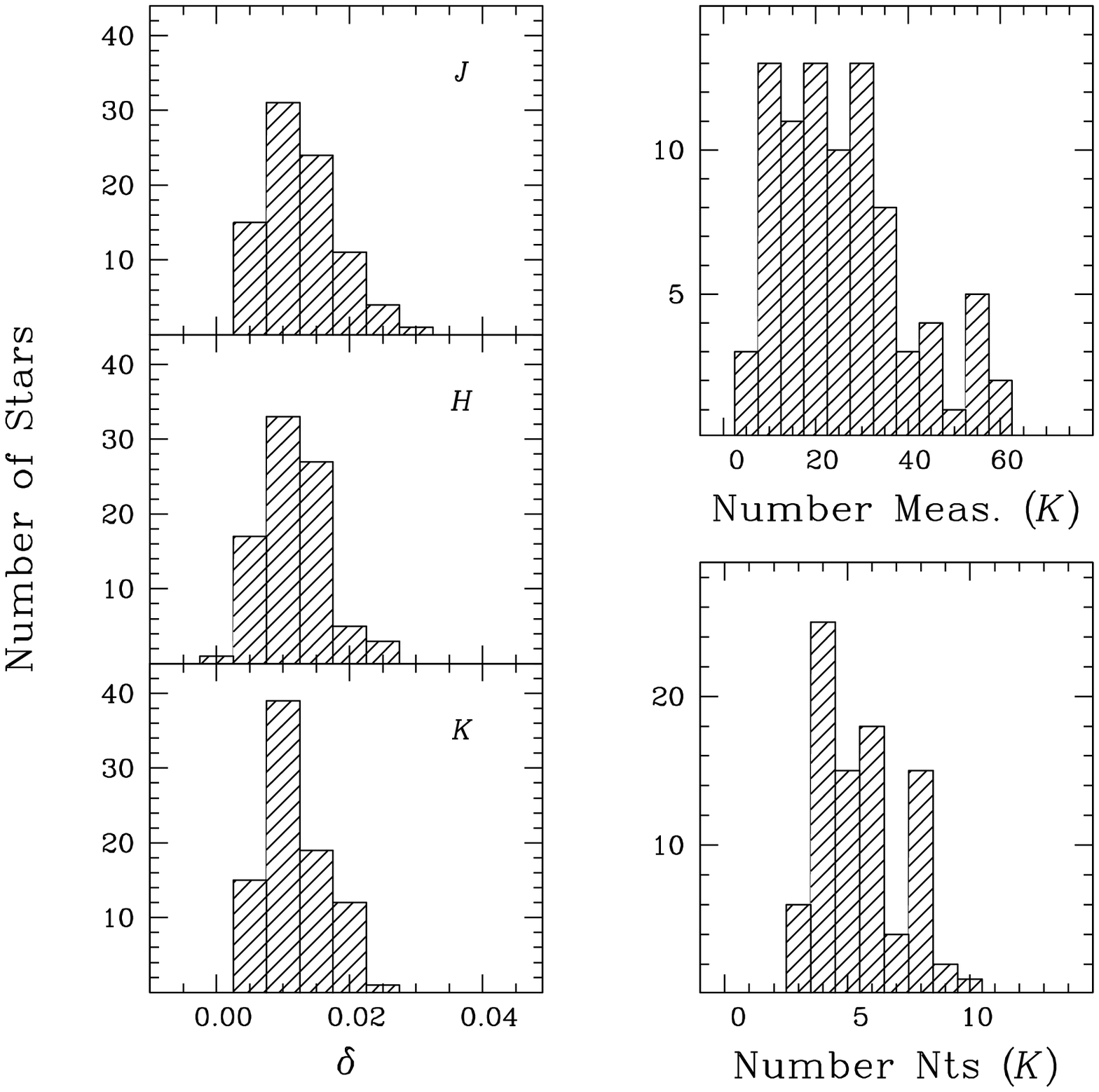}
\figcaption[fig2.ps]{
Histograms of 
$J$, $H$, and $K$ errors for the photometry reported in Table \ref{table:dat},
number of measurements (frames), and number of different nights.
Numbers of measurements and nights are shown for the $K$ band only.
The calculation of the errors of the mean $\delta$ shown in the left panels is described
in the Appendix.
\label{fig:stats} }
\end{figure}

The final database contains 6551 photometric measurements
from 3899 reduced images.
On average, stars have been observed on 26.0 times per filter on 5.5 different
nights, but a few have been observed on only three nights (see Table \ref{table:dat}), 
and these should probably be used with more discretion than the rest.
The median (mean) error in the photometry in Table~\ref{table:dat}
is 0\fm012 (0\fm013), 0\fm011 (0\fm012), and 0\fm011 (0\fm012), 
in $J$, $H$, and $K$, respectively. 
No photometric value in the Table has a formal error $>$\,0\fm028, and
standard deviations of the ensemble of values used
to calculate the photometry are always $\le$\,0\fm05.
Figure \ref{fig:stats} illustrates this quality assessment.
For a given star,
the measurements taken in the same night are not considered as 
statistically independent values, since
they were calibrated with the same zp and, in a given
five-position set, reduced with inter-dependent sky frames.

We note that the filling factor of the NICMOS3 array is roughly 95\%.
The projected pixel dimension at all telescopes was typically half the typical
seeing width, so that
the mean photometric error due to an incompletely filled array element 
is $\lesssim$\,0.1\% (McCaughrean \cite{mccaugh}), and should be negligible 
relative to the other sources of scatter.

\begin{figure}
\plotone{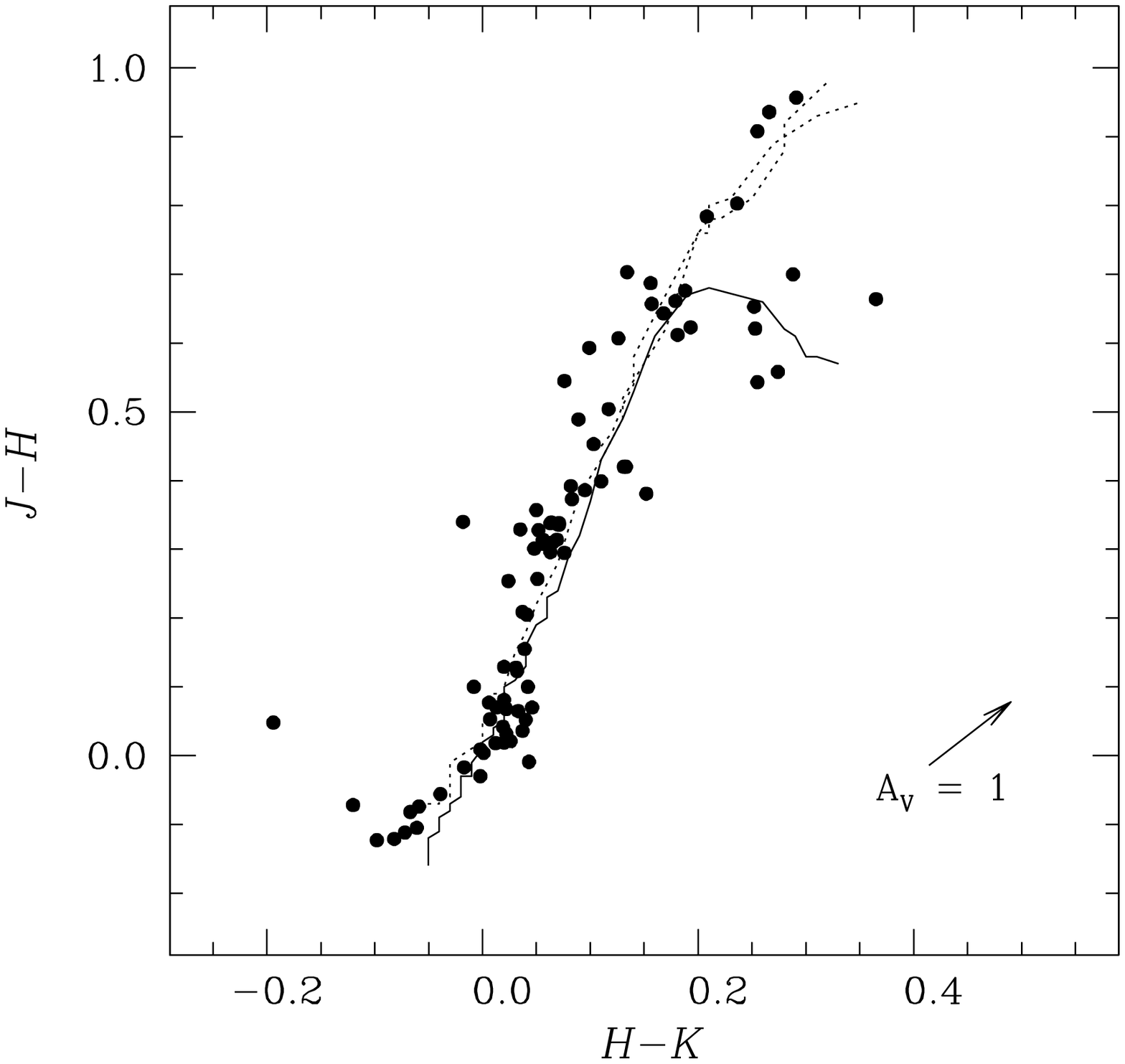}
\figcaption[color.ps]{$J-H$ vs. $H-K$ colors for the stars listed
in Table 1.
The solid line traces the main sequence (Koornneef \cite{koor}), and
the dotted lines the giants and supergiants.
A unit $V$-magnitude extinction is shown by the arrow in the lower right-hand
corner.
The isolated point with $J-H\,\sim\,0$ and $H-K\,\sim\,-0.2$ is FS~14.
\label{fig:color} }
\end{figure}

The $J-H$ and $H-K$ colors for the stars listed in Table~\ref{table:dat} 
are shown in Fig.~\ref{fig:color}.
It can be seen from the figure that the stars in our data set span a
wide range in spectral type, notwithstanding the initial selection
of a preponderance of A stars.
The $J-K$ color ranges from $-0.2$ to $\gtrsim\,1.2$,
making possible accurate evaluations of color transformations
between this photometry and other work.

We have checked for correlations of reported errors with  $J-K$ color or magnitude,
and find no trend with color.
The only (very weak) correlation is with $K$-band errors and $K$ magnitude, such
that errors are larger for fainter stars.
This is not unexpected behavior,
since the faintest stars suffer the most from the large
$K$-band background (typically 12--12.5\,mag\,arcsec$^{-2}$), 
and the consequent uncertainty in its subtraction.
Altough our data are well-suited to searches for variability, given the long 
month/years timescales spanned by our observations,
we found no convincing evidence for variation.
Nevertheless, a more definitive statement awaits long-term monitoring, in particular, 
of the reddest stars. 

\begin{figure}
\plotone{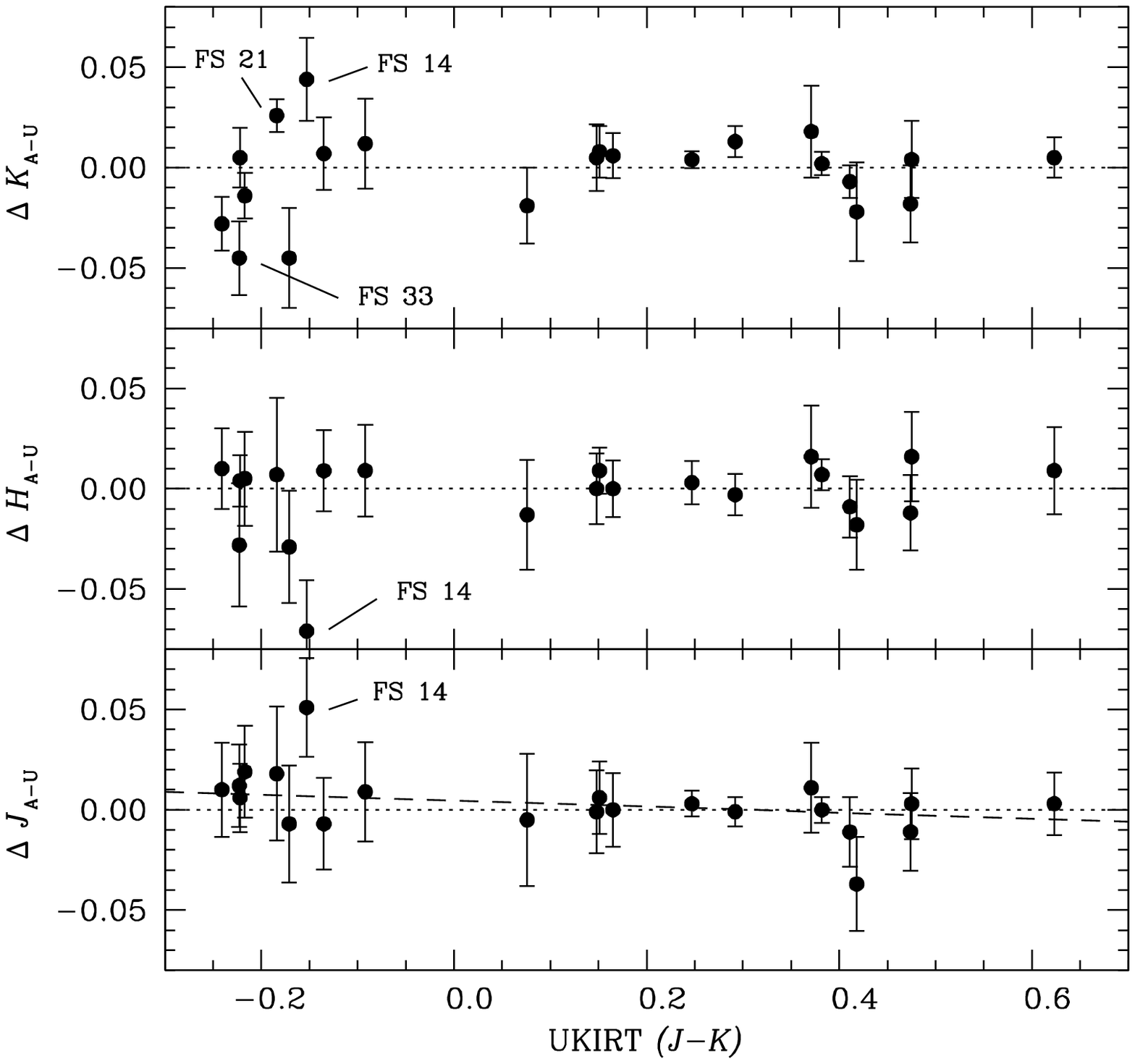}
\figcaption[fig4.ps]{
$J$-, $H$-, and $K$-band residuals shown as a function of UKIRT $J-K$;
the sense is ARNICA (this paper) - UKIRT.
Error bars are the quadrature sum of the errors associated with the original
UKIRT photometry (Casali \& Hawarden \cite{casa}), 
and the errors shown in Table \ref{table:dat}.
Residuals constant with $J-K$ color are shown by dotted lines, and the lower
panel shows the weighted regression (dashed line) 
of $J$-band residuals with $J-K$ color. 
Also indicated are the three stars with significant residuals as described in the text.
\label{fig:trans} }
\end{figure}

\subsection{Comparison with UKIRT Photometry and Transformation Equations }

Of the 86 stars in Table \ref{table:dat}, 22 are UKIRT Faint Standards 
(denoted as ``FS'' in the Table), and were used to calibrate the photometry.
The residuals for these stars were defined, similarly to the magnitudes
of the program stars, as the mean difference, over all the nights of observation,
between the nightly zp and the zp of the star itself.  
Their magnitudes, in the system described here, were determined from the sum
of the nominal UKIRT magnitudes and the mean residuals.
The mean residuals of the UKIRT faint standards (omitting FS~14,
see below)
are 0.001, -0.0004, and -0.004 in $J$, $H$, and $K$, respectively,
and with RMS differences of 0.012, 0.013, and 0.019.
We therefore measure no significant offset between the ARNICA photometry
reported here, and the UKIRT system from which it was derived.

Of the 22 UKIRT Faint Standards, three stars, FS~14, FS~21, and FS~33, have significant 
non-zero residuals relative to the published UKIRT photometry
(Casali \& Hawarden \cite{casa}).
The star with the largest residual ($\Delta\,H\,$=\,-0\fm061 with S/N\,=\,2.5), 
FS~14, is also the faintest star used as a calibrator.
Hence, it was subsequently treated as a program object and removed from the list 
of calibrators;
for all of the nights in which it was observed, the nightly calibration was redone,
and the photometry recalibrated.

We have attempted to derive a color transformation equation
between the photometry published here and the original UKIRT system. 
(A transformation equation between UKIRT and CIT has been published in
Casali \& Hawarden (\cite{casa}).)
The ARNICA $J$-, $H$-, and $K$-band residuals (in the sense ARNICA\,$-$\,UKIRT)
for the UKIRT calibrator stars and UKIRT $J-K$ color were fit with 
weighted regressions. 
(FS~14 was omitted from this operation.)
The only (possibly) significant non-zero slope is for $J$ with
$\Delta \,J = -0.015\,(J-K)_{UKIRT} + 0.0046$;
this regression is shown as a dashed line
in the lower panel of Fig. \ref{fig:trans}.
Although the ARNICA $J$ filter may be slightly redder than the UKIRT one,
as suggested by the above equation,
the data used to determine the regression are not suited to this kind of analysis.
The color excursion of the UKIRT FS's is limited ($J-K\,\sim\,-0.02\,-\,+0.6$),
and there is a trend between color and magnitude such that the bluest stars are also the
faintest, and thus subject to the largest uncertainty.
We caution, therefore, that these values for a $J$-band color transformation 
should be considered as very preliminary.

\section{Conclusions and Summary}

We have presented new NIR photometry for 86 stars in 40 fields.
The sky coverage is relatively uniform, and ideal for observatories
with $\delta\,\gtrsim\,30^\circ$.
On average,
stars have been observed on more than five different nights, and
typical errors on the photometry are 0\fm012.
We find some indication of a color transformation between the ARNICA 
(NICMOS) photometry reported here and the original UKIRT (InSb) system, 
but a definitive statement awaits a larger data set designed specifically 
to determine such a transformation.

\acknowledgments

The staffs of the TIRGO, the NOT, and the VATT helped make these observations
possible.
We are grateful to Colin Aspin, Mark Casali, and Tim Hawarden for insightful
comments and useful advice;
to Prof. Gianni Tofani for his generosity in funding the expenses of publication;
and to an anonymous referee for helpful criticism and comments on an earlier
version of this paper. 

\clearpage

\null
\bigskip

\appendix
\section{Calculation of the Means and Errors in Table \ref{table:dat}}

There are two main sources of error or uncertainty in our data:
residual spatial variation in the photometry (that is to say after the
flat-field correction), and variation in the atmospheric transparency.
In order for our error estimate to be reliable, we must be able to separate these
effects, and realistically assess their impact. 
We note that residual spatial variation is observed only with the optics used at 
the TIRGO.
The largest-amplitude effect was observed in the commissioning runs
(1992-93), and was corrected for by an empirically-determined flat field
mentioned in the main text, and described in detail in Hunt et al. (\cite{hunt96}).
The TIRGO optical train was then modified, and typical subsequent runs at TIRGO 
are characterized by roughly 0.05~mag scatter, 
a value which comprises both atmospheric and spatial variation.
Runs at the VATT and the NOT do not show measurable spatial variation in the photometry.

We describe here how the individual measurements were combined into the
final mean magnitudes reported in Table \ref{table:dat}.
The nightly magnitudes for each star and their associated errors are described first, followed
by the description of the final averages. 
As noted in the main text, in each filter,
a single observation consists of at least five frames (``measurements'') with
the star placed in different positions on the array.
In what follows, $\sigma$ will be used to denote exclusively the
standard deviation of a given ensemble.
The {\it errors} we associate with a given value will be denoted by $\delta$.

First, the variation in the atmospheric transparency $\sigma_t$ for each night 
was determined by the global fit, over an entire run, of the extinction 
coefficients and nightly zero points (zp) at unit air mass.
The photometry used for this fit was averaged over position, and thus, as long
as this averaging procedure is free of systematics, should be independent
of spatial variation.
The fit gives nightly zp's with formal errors, and these last define $\sigma_t$.
 
The residual spatial variation in the photometry was managed by using, 
when possible, positional zp's ($zp_p$) to separately
calibrate the program star measurements at each position. 
Such a procedure does not change the nightly mean, but reduces the impact of any
spatial variation in the photometry, which is
at least partially compensated by the equivalent spatial variation in the calibrator.
If on a given night $n$ we have $I_n$ measurements $f_*(p,n)$ of a given star, 
then we can define $m_{p,n} = -2.5\,\log\,f_*(p,n) + zp_{p,n}$,
where subscript $p$ indicates the position on the array.
The nightly mean magnitude $m_n$ was determined by a simple average over the $I_n$
measurements $m_{p,n}$.
Typically, $I_n$ is 5 for central stars, and less for field stars 
sufficiently distant from the central one.

The variance on the ensemble $m_{p,n}$ due to any residual spatial variation can be written as:
$$\sigma^2_p = \sigma^2_{x_p} + \sigma^2_{zp_p} + 2\,\sigma_{x_p,zp_p} \quad ,$$
where we have defined $x_{p,n} \equiv -2.5\,\log\,f_*(p,n)$;
$\sigma^2_{x_p}$ and $\sigma^2_{zp_p}$ are the spatial variances 
of the $x_p$ and $zp_p$, respectively, 
and $\sigma_{x_p,zp_p}$ is the covariance of the $x_p$ and $zp_p$.
If the photometry varies spatially, then $x_p$ and $zp_p$ are 
anti-correlated, and the covariance term {\it reduces} $\sigma_p$ accordingly.
If the use of positionally-dependent zp's is not possible 
(for example in the case of the field stars), 
then $\sigma^2_p$ takes into account
the increased scatter introduced by any spatial variation.

The final nightly error $\delta_n$ associated with $m_n$ was defined as the quadrature sum of 
the transparency variation $\sigma_t$, the uncertainty on the zp given by
$\sigma_t/N_{FS}$ where $N_{FS}$ is the number of calibrators observed on the $n^{th}$ night,
and the residual spatial variation (error on the mean):
$$\delta_n = \sqrt{\sigma_t^2 + \sigma_t^2/N_{FS} + \sigma_p^2/I_n} \quad .$$

The global mean of a given star reported in Table \ref{table:dat} is simply the 
weighted average $\langle m \rangle$ of $m_n$ over all the nights of observations, where
the weights are given by $1/\delta_n^2$. 
The error $\delta$ we associate with the final photometry is the error of the mean,
defined relative to the weighted average (as opposed to
the simple average which would minimize, by definition, the scatter).
We chose $\delta$ instead of the error of the weighted mean
because the former is proportional to the observed final scatter of the data, 
while the latter is determined exclusively by our estimate of the nightly 
errors.

\clearpage

\begin{deluxetable}{llrrrrrrrc} 
\tablecolumns{10} 
\tableheadfrac{0.05} 
\small
\tablewidth{0pt}
\tablenum{1}
\tablecaption{ARNICA Standard Star Photometry \label{table:dat}} 
\tablehead{ 
\multicolumn{1}{c}{Star } & \multicolumn{1}{c}{Other} & 
\multicolumn{2}{c}{Coordinates (2000)} &  &&&&& \multicolumn{1}{c}{Number} \\ 
\colhead{Name} &  \colhead{Designation} & 
\colhead{$\alpha$} & \colhead{$\delta$} & 
\colhead{$J$ ($\delta_J$)} & 
\colhead{$H$ ($\delta_H$)} & 
\colhead{$K$ ($\delta_K$)} & 
\colhead{$J-K$} & \colhead{$H-K$} & \colhead{Nights}\tablenotemark{a}} 
\startdata 
AS01-0 & FS02          & 00 55 09.9 &  00 43 13 & 10.716 (0.005) & 10.507 (0.010) & 10.470 (0.003) &  0.25 &  0.04 & 6,6,6 \nl
\tablevspace{6pt}
AS02-0 & SAO054271     & 00 55 58.6 &  39 10 09 &  8.775 (0.006) &  8.771 (0.012) &  8.770 (0.009) &  0.01 &  0.00 & 5,6,6 \nl
\tablevspace{6pt}
AS03-0 & FS03          & 01 04 21.6 &  04 13 39 & 12.606 (0.011) & 12.729 (0.008) & 12.827 (0.013) & -0.22 & -0.10 & 6,6,6 \nl
\tablevspace{6pt}
AS04-0 & FS04          & 01 54 37.6 &  00 43 01 & 10.555 (0.004) & 10.301 (0.005) & 10.277 (0.006) &  0.28 &  0.02 & 5,5,5 \nl
AS04-1 &               & 01 54 43.4 &  00 43 59 & 12.371 (0.021) & 12.033 (0.020) & 11.962 (0.025) &  0.41 &  0.07 & 4,3,4 \nl
\tablevspace{6pt}
AS05-0 & FS06          & 02 30 16.4 &  05 15 52 & 13.232 (0.009) & 13.314 (0.007) & 13.381 (0.010) & -0.15 & -0.07 & 8,7,8 \nl
AS05-1 &               & 02 30 18.6 &  05 16 42 & 14.350 (0.013) & 13.663 (0.011) & 13.507 (0.009) &  0.84 &  0.16 & 7,8,8 \nl
\tablevspace{6pt}
AS06-0 & SAO038218     & 02 41 03.6 &  47 41 28 &  8.713 (0.010) &  8.694 (0.014) &  8.674 (0.010) &  0.04 &  0.02 & 8,8,8 \nl
\tablevspace{6pt}
AS07-0 & FS07          & 02 57 21.2 &  00 18 39 & 11.105 (0.013) & 10.977 (0.009) & 10.946 (0.010) &  0.16 &  0.03 & 6,6,7 \nl
\tablevspace{6pt}
AS08-0 & SAO056596     & 03 38 08.3 &  35 10 52 &  8.744 (0.011) &  8.723 (0.014) &  8.697 (0.012) &  0.05 &  0.03 & 5,5,6 \nl
AS08-1 &               & 03 38 12.0 &  35 10 11 &  8.772 (0.011) &  7.836 (0.018) &  7.570 (0.017) &  1.20 &  0.27 & 5,5,5 \nl
AS08-2 &               & 03 38 08.3 &  35 09 38 &  9.728 (0.011) &  9.121 (0.015) &  8.995 (0.016) &  0.73 &  0.13 & 5,5,6 \nl
\tablevspace{6pt}
AS09-0 & SAO013053     & 04 11 05.8 &  60 10 21 &  8.432 (0.012) &  8.380 (0.014) &  8.340 (0.011) &  0.09 &  0.04 & 6,6,6 \nl
AS09-1 &               & 04 11 06.2 &  60 09 39 & 11.546 (0.015) & 10.923 (0.016) & 10.730 (0.015) &  0.82 &  0.19 & 6,6,6 \nl
\tablevspace{6pt}
AS10-0 & FS11          & 04 52 58.9 & -00 14 41 & 11.349 (0.011) & 11.281 (0.009) & 11.259 (0.006) &  0.09 &  0.02 & 8,8,7 \nl
\tablevspace{6pt}
AS11-0 & SAO058110     & 05 29 55.5 &  39 38 59 &  9.151 (0.009) &  9.181 (0.010) &  9.183 (0.011) & -0.03 & -0.00 & 7,8,8 \nl
AS11-1 &               & 05 30 02.0 &  39 37 49 & 11.299 (0.009) & 10.342 (0.015) & 10.051 (0.008) &  1.25 &  0.29 & 7,6,6 \nl
\tablevspace{6pt}
AS12-0 & FS12          & 05 52 27.4 &  15 53 23 & 13.700 (0.018) & 13.812 (0.015) & 13.884 (0.011) & -0.18 & -0.07 & 5,4,3 \nl
AS12-1 &               & 05 52 21.5 &  15 52 44 & 11.241 (0.007) & 10.931 (0.018) & 10.866 (0.020) &  0.38 &  0.06 & 3,4,4 \nl
\tablevspace{6pt}
AS13-0 & FS13          & 05 57 07.5 &  00 01 11 & 10.517 (0.005) & 10.189 (0.005) & 10.137 (0.005) &  0.38 &  0.05 & 9,9,8 \nl
AS13-1 &               & 05 57 10.4 &  00 01 38 & 12.201 (0.019) & 11.781 (0.007) & 11.648 (0.009) &  0.55 &  0.13 & 5,5,5 \nl
AS13-2 &               & 05 57 09.5 &  00 01 50 & 12.521 (0.008) & 12.101 (0.005) & 11.970 (0.006) &  0.55 &  0.13 & 5,5,5 \nl
AS13-3 &               & 05 57 08.0 &  00 00 07 & 13.345 (0.015) & 12.964 (0.017) & 12.812 (0.012) &  0.53 &  0.15 & 5,4,4 \nl
\tablevspace{6pt}
AS14-0 & SAO013747     & 06 11 25.4 &  61 32 15 &  8.688 (0.013) &  8.623 (0.014) &  8.590 (0.007) &  0.10 &  0.03 & 6,6,6 \nl
AS14-1 &               & 06 11 29.9 &  61 32 05 &  9.871 (0.015) &  9.801 (0.015) &  9.755 (0.011) &  0.12 &  0.05 & 6,6,6 \nl
\tablevspace{6pt}
AS15-0 & NGC2264       & 06 40 34.3 &  09 19 13 & 10.874 (0.007) & 10.669 (0.010) & 10.628 (0.008) &  0.25 &  0.04 & 8,9,10 \nl
AS15-1 &               & 06 40 36.2 &  09 18 60 & 12.656 (0.011) & 11.980 (0.010) & 11.792 (0.011) &  0.86 &  0.19 & 9,9,8 \nl
AS15-2 &               & 06 40 37.9 &  09 18 41 & 13.711 (0.014) & 12.927 (0.017) & 12.719 (0.012) &  0.99 &  0.21 & 8,9,8 \nl
AS15-3 &               & 06 40 37.9 &  09 18 19 & 14.320 (0.015) & 13.667 (0.017) & 13.415 (0.015) &  0.91 &  0.25 & 8,7,8 \nl
\tablevspace{6pt}
AS16-0 & FS14          & 07 24 14.3 & -00 33 05 & 14.159 (0.021) & 14.111 (0.011) & 14.305 (0.017) & -0.15 & -0.19 & 5,4,4 \nl
AS16-1 &               & 07 24 13.1 & -00 32 54 & 13.761 (0.020) & 13.638 (0.005) & 13.606 (0.021) &  0.15 &  0.03 & 5,4,4 \nl
AS16-2 &               & 07 24 15.4 & -00 32 49 & 11.411 (0.010) & 11.428 (0.017) & 11.445 (0.008) & -0.03 & -0.02 & 5,4,4 \nl
AS16-3 &               & 07 24 17.2 & -00 32 27 & 13.891 (0.012) & 13.855 (0.011) & 13.818 (0.011) &  0.07 &  0.04 & 5,4,3 \nl
AS16-4 &               & 07 24 17.5 & -00 33 07 & 11.402 (0.015) & 11.106 (0.019) & 11.043 (0.009) &  0.36 &  0.06 & 4,4,4 \nl
\tablevspace{6pt}
AS17-0 & NGC2419       & 07 38 15.5 &  38 56 16 & 14.353 (0.012) & 14.039 (0.015) & 13.983 (0.016) &  0.37 &  0.06 & 6,5,5 \nl
AS17-1 &               & 07 38 19.1 &  38 55 17 & 12.434 (0.009) & 12.077 (0.012) & 12.027 (0.017) &  0.41 &  0.05 & 6,6,6 \nl
AS17-2 &               & 07 38 16.6 &  38 57 13 & 14.745 (0.012) & 14.124 (0.012) & 13.871 (0.016) &  0.87 &  0.25 & 5,5,3 \nl
AS17-3 &               & 07 38 10.7 &  38 57 11 & 13.077 (0.007) & 12.782 (0.010) & 12.706 (0.006) &  0.37 &  0.08 & 6,6,4 \nl
AS17-4 &               & 07 38 08.2 &  38 56 01 & 14.459 (0.018) & 13.795 (0.014) & 13.430 (0.018) &  1.03 &  0.37 & 5,5,5 \nl
\tablevspace{6pt}
AS18-0 & FS15          & 08 51 05.8 &  11 43 47 & 12.741 (0.008) & 12.402 (0.004) & 12.338 (0.013) &  0.40 &  0.06 & 8,8,8 \nl
AS18-1 &               & 08 51 03.5 &  11 45 03 & 10.769 (0.024) & 10.614 (0.014) & 10.575 (0.014) &  0.19 &  0.04 & 4,6,6 \nl
\tablevspace{6pt}
AS19-0 & FS17          & 08 51 19.7 &  11 52 11 & 12.670 (0.014) & 12.334 (0.013) & 12.263 (0.004) &  0.41 &  0.07 & 5,5,6 \nl
AS19-1 &               & 08 51 21.8 &  11 52 38 & 10.095 (0.017) &  9.794 (0.013) &  9.746 (0.010) &  0.35 &  0.05 & 4,4,5 \nl
AS19-2 &               & 08 51 20.2 &  11 52 48 & 12.745 (0.019) & 12.488 (0.024) & 12.437 (0.012) &  0.31 &  0.05 & 4,4,5 \nl
\tablevspace{6pt}
AS20-0 & SAO042804     & 09 19 27.5 &  43 31 46 &  9.551 (0.012) &  9.519 (0.007) &  9.497 (0.015) &  0.05 &  0.02 & 9,9,9 \nl
\tablevspace{6pt}
\hline
\tablebreak
AS21-0 & SAO062058     & 10 28 42.1 &  36 46 18 &  9.061 (0.007) &  9.043 (0.015) &  9.031 (0.007) &  0.03 &  0.01 & 6,7,9 \nl
\tablevspace{6pt}
AS22-0 & FS21          & 11 37 05.6 &  29 47 59 & 12.966 (0.005) & 13.038 (0.010) & 13.158 (0.007) & -0.19 & -0.12 & 7,7,8 \nl
\tablevspace{6pt}
AS23-0 & SAO119183     & 12 02 53.5 &  04 08 47 &  8.889 (0.010) &  8.847 (0.009) &  8.828 (0.010) &  0.06 &  0.02 & 6,7,6 \nl
\tablevspace{6pt}
AS24-0 & SAO015832     & 12 35 15.6 &  62 56 57 &  9.398 (0.009) &  9.345 (0.009) &  9.338 (0.010) &  0.06 &  0.01 & 5,6,6 \nl
\tablevspace{6pt}
AS25-0 & FS33          & 12 57 02.4 &  22 02 01 & 14.029 (0.008) & 14.134 (0.010) & 14.195 (0.009) & -0.17 & -0.06 & 5,4,5 \nl
\tablevspace{6pt}
AS26-0 & FS23          & 13 41 43.6 &  28 29 51 & 13.000 (0.015) & 12.455 (0.012) & 12.379 (0.010) &  0.62 &  0.08 & 7,7,8 \nl
AS26-1 &               & 13 41 47.2 &  28 29 49 & 12.120 (0.016) & 11.527 (0.016) & 11.428 (0.016) &  0.69 &  0.10 & 7,8,8 \nl
\tablevspace{6pt}
AS27-0 & FS24          & 14 40 07.0 &  00 01 45 & 10.910 (0.015) & 10.781 (0.007) & 10.761 (0.010) &  0.15 &  0.02 & 5,4,5 \nl
AS27-1 &               & 14 40 07.3 &  00 02 23 & 12.991 (0.015) & 12.677 (0.024) & 12.608 (0.019) &  0.38 &  0.07 & 4,3,4 \nl
\tablevspace{6pt}
AS28-0 & SAO064525     & 15 09 20.3 &  39 25 49 &  8.455 (0.010) &  8.378 (0.014) &  8.372 (0.008) &  0.08 &  0.01 & 7,7,8 \nl
\tablevspace{6pt}
AS29-0 & FS25          & 15 38 33.3 &  00 14 19 & 10.234 (0.005) &  9.842 (0.013) &  9.760 (0.009) &  0.47 &  0.08 & 5,5,5 \nl
AS29-1 &               & 15 38 30.6 &  00 14 21 & 13.904 (0.017) & 13.566 (0.019) & 13.503 (0.020) &  0.40 &  0.06 & 5,3,4 \nl
\tablevspace{6pt}
AS30-0 & FS27          & 16 40 41.6 &  36 21 13 & 13.505 (0.004) & 13.197 (0.011) & 13.141 (0.014) &  0.36 &  0.06 & 4,4,4 \nl
AS30-1 &               & 16 40 36.6 &  36 22 40 & 12.497 (0.017) & 12.157 (0.011) & 12.175 (0.021) &  0.32 & -0.02 & 3,3,3 \nl
\tablevspace{6pt}
AS31-0 & FS28          & 17 44 06.8 & -00 24 58 & 10.744 (0.008) & 10.644 (0.005) & 10.602 (0.004) &  0.14 &  0.04 & 6,6,6 \nl
AS31-1 &               & 17 44 06.2 & -00 24 22 & 12.504 (0.010) & 12.131 (0.011) & 12.048 (0.007) &  0.46 &  0.08 & 6,6,6 \nl
AS31-2 &               & 17 44 04.9 & -00 24 11 & 13.290 (0.009) & 12.633 (0.011) & 12.476 (0.010) &  0.81 &  0.16 & 5,5,4 \nl
\tablevspace{6pt}
AS32-0 & SAO017946     & 18 36 10.5 &  65 04 30 &  8.017 (0.018) &  7.917 (0.012) &  7.925 (0.014) &  0.09 & -0.01 & 4,4,4 \nl
\tablevspace{6pt}
AS33-0 & FS35          & 18 27 13.6 &  04 03 10 & 12.220 (0.004) & 11.834 (0.005) & 11.739 (0.009) &  0.48 &  0.09 & 4,4,5 \nl
AS33-1 &               & 18 27 12.4 &  04 02 16 & 13.180 (0.009) & 12.477 (0.004) & 12.343 (0.008) &  0.84 &  0.13 & 4,4,5 \nl
AS33-2 &               & 18 27 15.5 &  04 03 34 & 13.724 (0.024) & 13.271 (0.023) & 13.168 (0.014) &  0.56 &  0.10 & 4,4,5 \nl
\tablevspace{6pt}
AS34-0 & SAO048300     & 19 17 34.5 &  48 06 02 &  8.443 (0.019) &  8.434 (0.003) &  8.436 (0.019) &  0.01 & -0.00 & 3,3,3 \nl
\tablevspace{6pt}
AS35-0 & SAO070237     & 20 35 11.2 &  36 08 49 &  8.625 (0.009) &  8.544 (0.011) &  8.524 (0.006) &  0.10 &  0.02 & 3,4,4 \nl
\tablevspace{6pt}
AS36-0\tablenotemark{b} & FS29          & 21 52 25.4 &  02 23 20  & 13.168 (0.013) & 13.242 (0.007) & 13.301 (0.007) & -0.13 & -0.06 & 4,4,4 \nl
AS36-1 &               & 21 52 26.2 &  02 24 41 & 13.538 (0.028) & 13.049 (0.012) & 12.960 (0.014) &  0.58 &  0.09 & 3,4,4 \nl
AS36-2 &               & 21 52 22.4 &  02 24 32 & 14.246 (0.014) & 13.847 (0.014) & 13.737 (0.008) &  0.51 &  0.11 & 4,4,4 \nl
AS36-3 &               & 21 52 21.8 &  02 22 51 & 14.165 (0.004) & 13.607 (0.009) & 13.333 (0.022) &  0.83 &  0.27 & 4,3,4 \nl
AS36-4 &               & 21 52 25.4 &  02 23 35 & 15.312 (0.008) & 14.769 (0.006) & 14.514 (0.022) &  0.80 &  0.26 & 4,4,4 \nl
\tablevspace{6pt}
AS37-0 & SAO072320     & 22 25 20.7 &  40 09 38 &  8.669 (0.017) &  8.678 (0.009) &  8.635 (0.009) &  0.03 &  0.04 & 5,5,5 \nl
AS37-1 &               & 22 25 19.8 &  40 08 06 & 11.020 (0.026) & 10.516 (0.014) & 10.399 (0.011) &  0.62 &  0.12 & 3,3,3 \nl
\tablevspace{6pt}
AS38-0 & FS30          & 22 41 44.7 &  01 12 36 & 11.932 (0.006) & 11.988 (0.009) & 12.027 (0.010) & -0.09 & -0.04 & 5,6,4 \nl
AS38-1 &               & 22 41 46.4 &  01 11 52 & 12.994 (0.004) & 12.657 (0.012) & 12.588 (0.020) &  0.41 &  0.07 & 4,5,4 \nl
AS38-2 &               & 22 41 50.2 &  01 12 43 & 11.355 (0.023) & 11.026 (0.014) & 10.991 (0.013) &  0.36 &  0.04 & 4,5,4 \nl
\tablevspace{6pt}
AS39-0 & FS31          & 23 12 21.2 &  10 47 06 & 13.808 (0.008) & 13.929 (0.002) & 14.011 (0.009) & -0.20 & -0.08 & 5,5,4 \nl
AS39-1 &               & 23 12 20.7 &  10 46 36 & 15.033 (0.022) & 14.333 (0.004) & 14.045 (0.018) &  0.99 &  0.29 & 4,4,4 \nl
\tablevspace{6pt}
AS40-0 & NGC7790       & 23 58 50.2 &  61 10 02 & 11.051 (0.013) & 10.408 (0.013) & 10.240 (0.013) &  0.81 &  0.17 & 7,8,8 \nl
AS40-1 &               & 23 58 43.2 &  61 10 26 & 11.900 (0.012) & 11.097 (0.011) & 10.861 (0.008) &  1.04 &  0.24 & 7,8,7 \nl
AS40-2 &               & 23 58 59.7 &  61 10 26 & 11.477 (0.018) & 10.569 (0.014) & 10.314 (0.018) &  1.16 &  0.26 & 4,4,4 \nl
AS40-3 &               & 23 58 57.5 &  61 09 46 & 10.578 (0.015) &  9.966 (0.011) &  9.785 (0.008) &  0.79 &  0.18 & 6,6,6 \nl
AS40-4 &               & 23 58 53.6 &  61 11 02 & 10.356 (0.015) &  9.695 (0.008) &  9.516 (0.007) &  0.84 &  0.18 & 7,7,7 \nl
AS40-5 &               & 23 58 43.2 &  61 09 42 &  9.488 (0.013) &  9.418 (0.007) &  9.405 (0.008) &  0.08 &  0.01 & 6,6,8 \nl
\enddata 
\tablenotetext{a}{The three numbers refer to the number of nights in each
band, $J$, $H$, and $K$, respectively.
The number of nights may differ between bands because of lack of measurements,
and may differ between stars of a given field because of bad pixels
or positioning of the field on the detector.} 
\tablenotetext{b}{This star, also known as G93-48,
has large proper motion: 23.0, -303.0 mas/yr.}
\end{deluxetable} 


\clearpage

\begin{thebibliography}{}

\bibitem[1983]{alle}{
Allen, D.A. \& Cragg, T.A. 1983, \mnras, 203, 777 
}

\bibitem[1991]{bouchet}{
Bouchet, P., Schmider, F.X., \& Manfroid, J. 1991, A\&ASS, 91, 409
}

\bibitem[1990]{carter90}{
Carter, B.S. 1990, \mnras, 242, 1
}

\bibitem[1995]{carter95}{
Carter, B.S. \& Meadows, V.S. 1995, \mnras, 276, 734
}

\bibitem[1992]{casa}{
Casali, M. M. \& T. G. Hawarden 1992, JCMT-UKIRT Newsletter, No. 4, 33
}

\bibitem[1985]{chri}{
Christian, C.A., Adams, M., Barnes, J.V., Butcher, H., Hayes, D.S.,
Mould, J.R., \& Siegel, M. 1985, \pasp, 97, 363
}

\bibitem[1964]{egge}{
Eggen,~O.~J. \&  Sandage,~A.~R. 1964, \apj, 140, 130
}

\bibitem[1982]{elia82}{
Elias,~J.~H., Frogel,~J.~A., Matthews,~K., \& Neugebauer,~G. 1982, \aj, 87, 1029
}

\bibitem[1981]{enge}{
Engels, D., Sherwood, W.A., Wamsteker, W., and Shultz, G.V. 1981, A\&ASS, 45, 5 
}

\bibitem[1978]{frog}{
Frogel, J.A., Persson, S.E., Aaronson, M., \& Matthews, K. 1978, \apj, 220, 75 }

\bibitem[1974]{glas}{
Glass, I.S. 1974, MNAS Sth. Afr., 33, 53, 71
}

\bibitem[1980]{jone80}{
Jones, T.J. \& Hyland, A.R. 1980, \mnras, 192, 359
}

\bibitem[1982]{jone82}{
Jones, T.J. \& Hyland, A.R. 1982, \mnras, 200, 509 
}

\bibitem[1994a]{hunta}{
Hunt, L., Maiolino, R., \& Moriondo, G. 1994a,
Arcetri Observatory Technical Report N.~2/94}

\bibitem[1994b]{huntb}{
Hunt, L., Maiolino, R., Moriondo, G. \& Testi, L. 1994b,
Arcetri Observatory Technical Report N.~3/94}

\bibitem[1994c]{huntc}{
Hunt, L., Testi, L., Borelli, S., Maiolino, R., \& Moriondo, G. 1994c,
Arcetri Observatory Technical Report N.~4/94}

\bibitem[1996]{hunt96}{
Hunt,~L.K., Lisi,~F., Testi,~L., Baffa,~C., Borelli,~S., Maiolino,~R.,
Moriondo,~G., \& Stanga,~R.M. 1996, A\&ASS, 115, 181}

\bibitem[1983]{koor}{
Koornneef, J. 1983, \aap, 128, 84
}

\bibitem[1983]{land}{
Landolt, A.U. 1983, \aj, 88, 439
}

\bibitem[1993]{lisi93}{
Lisi, F., Baffa, C., \& Hunt, L. 1993, in SPIE Vol. 1946, Infrared Detectors
and Instrumentation, ed. A.M. Fowler, 594}

\bibitem[1996]{lisi96}{
Lisi, F., Baffa, C., Biliotti, V., Bonaccini, D., Del Vecchio, C., Gennari,
S., Hunt, L., Marcucci, G., \& Stanga, R.M. 1996, \pasp, 108, 364 }

\bibitem[1988]{mccaugh}{
McCaughrean, M. J. 1988, Ph.D. Thesis, University of Edinburgh}

\bibitem[1990]{turn}{
Turnshek, D.A., Bohlin, R.C., Williamson, R.L. II, Lupie, O.L., \&
Koornneef, J. 1990, \aj, 99, 1243
}

\end{thebibliography}
\end{document}